\documentclass[preprint,11pt]{elsarticle}
\usepackage{graphicx}
\usepackage{amssymb,amsmath}
\usepackage{caption}
\usepackage{subcaption}
\usepackage{morefloats}
\usepackage{booktabs}
\usepackage{natbib}
\biboptions{numbers,sort&compress}
\usepackage[colorlinks=true,linkcolor=black,citecolor=blue,urlcolor=blue]{hyperref}
\usepackage{epstopdf}
\begin{document}

\begin{frontmatter}
\title{Study of bioheat transfer phase change during cryosurgery for an irregular tumor tissue using EFGM}

\author{R. Bhargava\corref{cor1}}
\ead{rbharfma@iitr.ac.in}
\author{H. Chandra\corref{cor2}}
\ead{harishchandraamu@gmail.com}

\address{Department of Mathematics, Indian Institute Technology of Roorkee, Roorkee-247667}

\begin{abstract}
Cryosurgery has been consistently used as an effective treatment to eradicate irregular tumor tissues. During this process, many difficulties occur such as intense cooling may also damage the neighboring normal tissues due to the release of large amount of cold from the cooling probe. In order to protect the normal tissues in the vicinity of target tumor tissues, coolant was released in a regulated manner accompanied with the nanoparticle to regulate the size and orientation of ice balls formed together with improved probe capacity. The phase change occurs in the target tumor tissues during cryosurgery treatment. The effective heat capacity method is used for simulation of phase change in bio-heat transfer equation to take into account the latent heat of phase transition. The bio-heat transfer equation is solved by using element free Galerkin method (EFGM) to simulate the phase change problem of biological tissues subject to nano cryosurgery. In this study, Murshed model with cylindrical nanoparticles is used for the high thermal conductivity of nanofluids as compared to Leong Model with the spherical nanoparticle. The important effects of the interfacial layer at the mushy region (i.e. liquid to the solid interface), size and concentration of nanoparticles are shown on the freezing process. This type of problem has applications in biomedical treatment such as drug delivery. Application of cryosurgery in bio-fluids used for drug delivery in cancer therapy can be made more efficient in the presence of nanoparticles (such as Iron oxide ($Fe_{3}O_{4}$), alumina ($Al_{2}O_{3}$) and gold ($Au$)).
\end{abstract}


\end{frontmatter}

\begin{table}
\begin{tabular}{l l }
{\bf Nomenclature} & \\
$Roman ~Symbols$               & \\
$c$                            &     Specific Heat capacity\\
$d_{p}$                        &     Diameter of nanoparticle\\
$k$                            &     Thermal conductivity\\
$t$                            &     Time\\
$T_{0}$                        &     Initial temperature\\
$T$                            &     Temperature\\
$Greek ~Symbols$               &          \\
$\alpha$                       &     Parameter used in time integration-schemes\\
$\bar\alpha$                   &     Penalty parameter\\
$\delta$                       &     Nanolayer to base fluid conductivity ratio\\
$\rho$                         &     Density\\
$\phi$                         &     Volume fraction of nanoparticle\\
$w$                            &     Perfusion rate\\
$Subscripts$                   &           \\
$b$                            &     Blood\\
$eff$                          &     Effective\\
$f$                            &     Basefluid\\
$l$                            &     Liquid (fluid)\\
$lr$                           &     Interfacial layer\\
$nf$                           &     Nanofluid\\
$s$                            &     Solid (particle)\\
\end{tabular}
\end{table}

\section{Introduction}
Cryosurgery (commonly known as cryotherapy) is the process of surgery in which probes are used to generate extremely low temperature in the targeted tumor tissue which helps in the annihilation of tumor cells. A successful treatment requires the high cooling efficiency of cryoprobe for destroying target tumor tissues. Yan et al. \cite{YL} have been studied on nanocryosurgery and its mechanisms for the enhanced cooling capacity of tumor tissues. After that Liu and Deng \cite{LD} discussed the nanoparticles thermal conductivity effects on the cooling process such as enhancement of ice nucleation, transport of water during cooling of a single cell. Some possible challenges and applications are illustrated when nanocryosurgery used with cryosurgery. Extreme cooling was carried out by using a cryoprobe whose cooling efficiency can be enhanced considerably via loading with nanoparticles with high thermal conductivity of nanofluids. Size and orientation of ice ball, as well as the rate of heat transfer, can be regulated and optimized by tumor injecting nanoparticles with high thermal conductivity to provide safe and efficacious surgical treatment in removing tumor tissues.
\par
During cryosurgery process, the formation of ice crystals was observed within the extracellular spaces when temperature lies in the freezing range. Due to an extremely low temperature ranging from $-4^{0}C$ to $-21^{0}C$, extracellular crystallization cell destruction occurs which was observed by Chua et al. \cite{CCH}. The ice crystal grows continuously as the freezing time is increased leading to loss of liquid water which in turn results in cell shrinkage followed by the destruction of cell membrane. Using a single probe method, Singh et al. \cite{SB} analyzed the phase change problem during cryosurgery treatment of a biological tissue which is a standard shape of the tumor with spherical nanoparticles. The applications of cryosurgery are found in the management of urology, cancer, gynecology, dermatology, ophthalmology, neurosurgery etc \cite{G, R}.
\par
Size and orientation of ice ball in a multi-probe cryosurgery were studied by Liu et al. \cite{LMWR}. The dual reciprocity boundary element method was used to solve the modeling of a multidimensional freezing problem during cryosurgery by Deng and Liu \cite{DL}. After that Li et al. \cite{LLTH} investigated the phase transition problem in tumor surgery using $\alpha$-FEM. Various heat transfer problems are considered during the phase change problems for both types single and multiple moving boundary \cite{JSC, SK}. To simulate phase change problems of biological tissues in situ are not discussed but it has considered the effects of blood perfusion and metabolic heat production in the unfrozen region. The study of tissue and arterial blood temperature was initiated by Pennes' \cite{P}. The aim of Pennes' investigation was to calculate the capability of heat transfer theory to compute the rate of heat generation locally in the tissue and volume of blood flow in surrounding local tissue.
\par
Biocompatible nanoparticles like magnetite ($Fe_3O_4$), alumina ($Al_2O_3$) and gold ($Au$) are exploited for their clinical application owing to their safe and negligible side effects, which was proposed by Zhang et al. \cite{ZCWL}. Moreover, the ice ball formation during cryosurgery can be effectively regulated by injecting them with nanoparticle on the targeted tissues, therefore rendering cryosurgery more efficient. Conclusively, the application of nanoparticles into target tissues enhances permeability as well as image contrast resulting in the improved image to carry out cryosurgical operation proposed by Wickline et al. \cite{WL}. Several models have been developed for effective thermal conductivity of nanofluids due to the different nanoparticle in base fluids by the researchers. The classical models of Maxwell \cite{M} and Hamilton-Crosser (HC) \cite{HC} were developed for thermal conductivity of nanofluids which increases with the low volume fraction of nanoparticles. These models have not considered the effect of particle size, volume fraction, and the interfacial layer at the particle/liquid interface \cite{KPCE, EPCK}. According to previous studies \cite{LYM, XX} including the effect of particle size and nanoparticles dispersion along with particle-particle collision due to thermal conductivity, the volume fraction of nanoparticles have considered $20 \%$ by Nayak et al.\cite{NBP}.
\par
In the recent years, a different type of numerical scheme has been used to solve the heat transfer problem of phase change or boundary moving problem. These numerical methods may be categorized into two groups: (1) front tracking and other is (2) fixed grid method. Using front tracking approach \cite{ISC, VST}, two different domains solid and liquid are used for which the moving boundary condition is given by the latent heat. This technique used for successive deforming grid and size and shape of the whole in-silico domain is changed where the coordinate is changed with each time step, requiring high computation time. One more limitation of the front tracking techniques is, it is not applicable for a material problem with a finite cooling interval \cite{VST}. To overcome this obstacle and to simulate the phase change problem fixed grid is used frequently.
\par
In fixed grid technique, there are various approaches to solve viz. finite difference method, finite element method and finite volume method depending on the discretization of the domain. These grid based numerical techniques are not applicable for large deformation problem. As FEM has some inherent which comes from integration that depends on the element mesh. But such type of problems can be easily solved using the mesh-free numerical technique. Therefore, the mesh-free technique has been used here which is an advanced numerical technique, applicable for complex geometry. The mesh-free method does not require a mesh. The details of the mesh-free method can be seen in \cite{SB, L, STE, S, SSP}.
\par
In the present paper, the phase change problem during cryosurgery treatment for an irregular tumor tissue with natural convection has been studied. We have included the effective thermal conducting model of nanofluid proposed by Murshed et al. \cite{MLY1}. In this model cylindrical nanoparticle is used for freezing purpose which has more freezing capacity as compared to the spherical nanoparticle. The effect of nanoparticle size, nanolayer to base fluid conductivity ratio and volume fraction of nanoparticle on cooling process has been shown. To resolve the bio-heat transfer equation Element Free Galerkin method (EFGM) has been used.
\section{Mathematical Formulation}
 The bio-heat transfer model proposed by Pennes \cite{P} is used to simulate the heat flow. The core feature of Pennes model is that it involves only single variable in the simulation. The $2D$ unsteady bio-heat transfer equation can be written in the form as:
\begin{align}
\rho c \frac{\partial T}{\partial t}=k \nabla^{2} T+\omega_{b}\rho_{b} c_{b} (T_{b}-T)+Q_{m}-Q \label{e1}
\end{align}
Where $\rho$, $c$ and $k$ represent the density, specific heat capacity and thermal conductivity of tissue respectively. $\omega_{b}$ is the blood perfusion rate, $T_{b}$ is the blood temperature, $Q_{m}$ is the volumetric heat source associated with the metabolism and `$-Q$' is the external heat source.
\par
The bio-heat transfer equation for nanofluid can be written as
\begin{align}
(\rho c)_{n f} \frac{\partial T}{\partial t}=k_{n f} \nabla^{2} T+\omega_{b}(\rho_{b} c_{b})_{n f} (T_{b}-T)+Q_{m}-Q \label{e2}
\end{align}
The outer boundary of the computational domain is kept as $37 ^{0}C$. The initial temperature of the entire domain is fixed same as body temperature i.e. $T_{0}=37 ^{0} C$. In the present study, shape of tumor tissue was considered irregular in the physical domain (part of the human body) in which a single probe with radius of $1mm$ is inserted at the centre of the tumor for cooling purpose. The physical domain, containing irregular shape of tumor tissue is maintained at a constant temperature as shown in figure 1.
\begin{figure}[h!]
\vspace{-0.10in}
\centerline{\includegraphics[height=7.0cm,width=8.0cm]{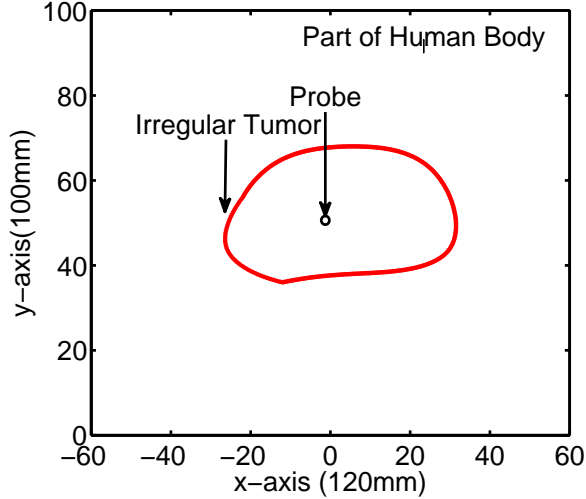}}
\vspace{-0.12in}
\caption{Physical domain of the problem.} \label{fig1}
\end{figure}
\subsection{The effective thermal conductivity model for nanofluid}
Nanoparticles enhance the thermal conductivity of nanofluid therefore the Murshed \cite{MLY1} model is used. The effective thermal conductivity $(k_{nf})$ for cylindrically shaped nanoparticles can be written as
\begin{align}
k_{n f}=\frac{(k_{p}-k_{lr})\phi k_{lr}[r^{2}_{2}-r^{2}_{1}+1]+(k_{p}+k_{lr})r^{2}_{2}[\phi_p r^{2}_{1}(k_{lr}-k_{f})+k_{f}]}{r^{2}_{2}(k_{p}+k_{lr})-(k_{p}-k_{lr})\phi_{p}[r^{2}_{2}+r^{2}_{1}-1]} \label{e3}
\end{align}
 where $1+\frac{h}{d_{p}}=r_{1}$ and $1+\frac{h}{2 d_p}=r_{2}$ \\
 Here $d_{p}$ is the radius of particle and $h=\sqrt{2 \pi \sigma}$ is the thickness of interfacial layer at the plane.
 \par
 where $\sigma$ is a parameter which represents the diffuseness of the interfacial boundary for cylindrical nanoparticles. It generally lies within $0.2-0.8$ nm. For $\sigma=0.4$, h was calculated as 1nm. Thermal conductivity of the interfacial layer thickness can be written as $k_{lr}=\delta k_{f}$.
 \par
 Where $\delta>1$, is an observational parameter depending on the distribution of fluid particle, if $\delta=1$, then there is no interfacial layer thickness at the solid(particle)/liquid interface by Leong et al. \cite{LYM}.\\
 and, simplified form of equation (\ref{e3}), can be written as:
\begin{align}
 k_{nf}=k_{f}\frac{\delta \phi (k_{p}-\delta k_{f})[r^{2}_{2}-r^{2}_{1}+1]+(k_{p}+\delta k_{f})r^{2}_{2}[\phi r^{2}_{1}(\delta-1)+1]}{r^{2}_{2}(k_{p}+\delta k_{f})-(k_{p}-\delta k_{f})\phi (r^{2}_{2}+r^{2}_{1}-1)} \label{e4}
 \end{align}
 The density of the nanofluid and heat capacitance of the nanofluid is given by:
 \begin{align}
  \rho_{n f}=(1-\phi)\rho_{f}+\phi \rho_{s} \nonumber \\
  (\rho c)_{n f}=(1-\phi)(\rho c)_{f}+\phi (\rho c)_{s} \label{e5}
  \end{align}
  where, $\phi$ represents volume fraction of the solid particles and subscripts $f$, $s$ and $nf$ denote base fluid, solid(particles) and nanofluid respectively.
  \subsection{Phase change model}
  The entire domain is divided into three different regions- frozen region, unfrozen region and mushy region. Initially, the whole computational domain contains only unfrozen region then after treatment with coolant probe it started freezing and attended the mushy form followed by frozen form.
  \par
  The following assumptions have been made in the simulation of phase transition of cryosurgery process \cite{CCH,SB,LLTH}:
\begin{itemize}
  \item Latent heat is constant and heat transfer varies at the point where, phase changes from liquid to solid.
  \item The heat transfer is purely conduction process at the interface.
  \item The mushy region is considered between the temperature range $T_{s}$ to $T_{l}$ where $T_{s} = -8^{0}C$ and $T_{l} = -1^{0}C$. After that the frozen region emerges.
  \item The density is considered as constant for solid and liquid phase.
  \item In the freezing phase, metabolism must be zero.
  \item The cylindrically shaped nanoparticles are loaded in the fluid.
  \end{itemize}
  Thus, bio-heat transfer equation can be described as
  \begin{align}
  (\rho c_{eff}) \frac{\partial T}{\partial t}=k(\frac{\partial^{2} T }{\partial x^{2}}+\frac{\partial^{2} T}{\partial y^{2}})+\omega_{b}(\rho_{b} c_{b}) (T_{b}-T)+Q_{m}-Q \label{e6}
 \end{align}
 Where, the effective heat capacity, thermal conductivity, metabolic heat generation and blood perfusion for different regions are as follows.
 \begin{align}
 (\rho c)_{eff} =
    \begin{cases}
            (\rho c)_{frozen},                                                          &          T<T_{s}\\
            \frac{(\rho c)_{frozen}+(\rho c)_{unfrozen}}{2}+\frac{\rho \bar h}{T_{l}-T_{s}}, &          T_{s}\leq T \leq T_{l} \label{e7}\\
            (\rho c)_{unfrozen},                                                        &          T>T_{l}
    \end{cases}
 \end{align}

 \begin{align}
 k =
    \begin{cases}
            (k)_{frozen},                                                          &          T<T_{s}\\
            \frac{(k)_{frozen}+(k)_{unfrozen}}{2},                                 &          T_{s}\leq T \leq T_{l} \label{e8}\\
            (k)_{unfrozen},                                                        &          T>T_{l}
    \end{cases}
 \end{align}

 \begin{align}
 Q_{m} =
    \begin{cases}
            {0},                                                                   &          T<T_{s}\\
            {0},                                                                   &          T_{s}\leq T \leq T_{l} \label{e9}\\
            Q_{u},                                                                 &          T>T_{l}
    \end{cases}
 \end{align}

 \begin{align}
 \omega_{b} =
    \begin{cases}
            {0},                                                                  &          T<T_{s}\\
            {0},                                                                  &          T_{s}\leq T \leq T_{l} \label{e10}\\
             \omega_{u},                                                          &          T>T_{l}
    \end{cases}
 \end{align}
 Thermal conductivity and specific heat capacity for two different parts (biological and nanoparticle) are used for frozen and unfrozen region, the equations (\ref{e4}) and (\ref{e5}) can be written as follows.
 \begin{align}
 (k_{nf})_{frozen}=(k_{f})_{frozen}\frac{\delta \phi (k_{p}-\delta (k_{f})_{frozen})[r^{2}_{2}-r^{2}_{1}+1]+(k_{p}+\delta (k_{f})_{frozen})r^{2}_{2}[\phi r^{2}_{1}(\delta-1)+1]}{r^{2}_{2}(k_{p}+\delta (k_{f})_{frozen})-(k_{p}-\delta (k_{f})_{frozen})\phi (r^{2}_{2}+r^{2}_{1}-1)} \label{e11}
 \end{align}
 \begin{align}
 (k_{nf})_{unfrozen}=(k_{f})_{unfrozen}\frac{\delta \phi (k_{p}-\delta (k_{f})_{unfrozen})[r^{2}_{2}-r^{2}_{1}+1]+(k_{p}+\delta (k_{f})_{unfrozen})r^{2}_{2}[\phi r^{2}_{1}(\delta-1)+1]}{r^{2}_{2}(k_{p}+\delta (k_{f})_{unfrozen})-(k_{p}-\delta (k_{f})_{unfrozen})\phi (r^{2}_{2}+r^{2}_{1}-1)} \label{e12}
 \end{align}
and
\begin{align}
(\rho c_{n f})_{frozen}=(1-\phi)(\rho c_f)_{frozen}+\phi (\rho c_{s})_{frozen} \label{e13}\\
(\rho c_{n f})_{unfrozen}=(1-\phi)(\rho c_f)_{unfrozen}+\phi (\rho c_{s})_{unfrozen} \label{e14}
\end{align}

\section{Numerical solution}
\subsection{Element Free Galerkin Method}
In this study, the element free Galerkin method (EFGM) is used for numerical computation of temperature fields. With the increasing size of the ice-ball with the time, the nodes are spread in its vicinity. The whole computational domain is discretized with the assistance of nodes as shown in figure 2. The distribution density of the nodes in its neighbouring region depends upon the size of the generated ice ball. Moving least square (MLS) interpolation functions are required to approximate the unknown function for the execution of EFGM. The MLS approximation consists three parts: (i) a weight function (ii) a basis function (iii) set of non-constant coefficients. A small nbd around any node is defined as a support domain and the weight function corresponding to that node is non-zero throughout the domain of influence.
\begin{figure}
\begin{subfigure}[b]{0.5\textwidth}
  \includegraphics[height=7cm,width=8.5cm]{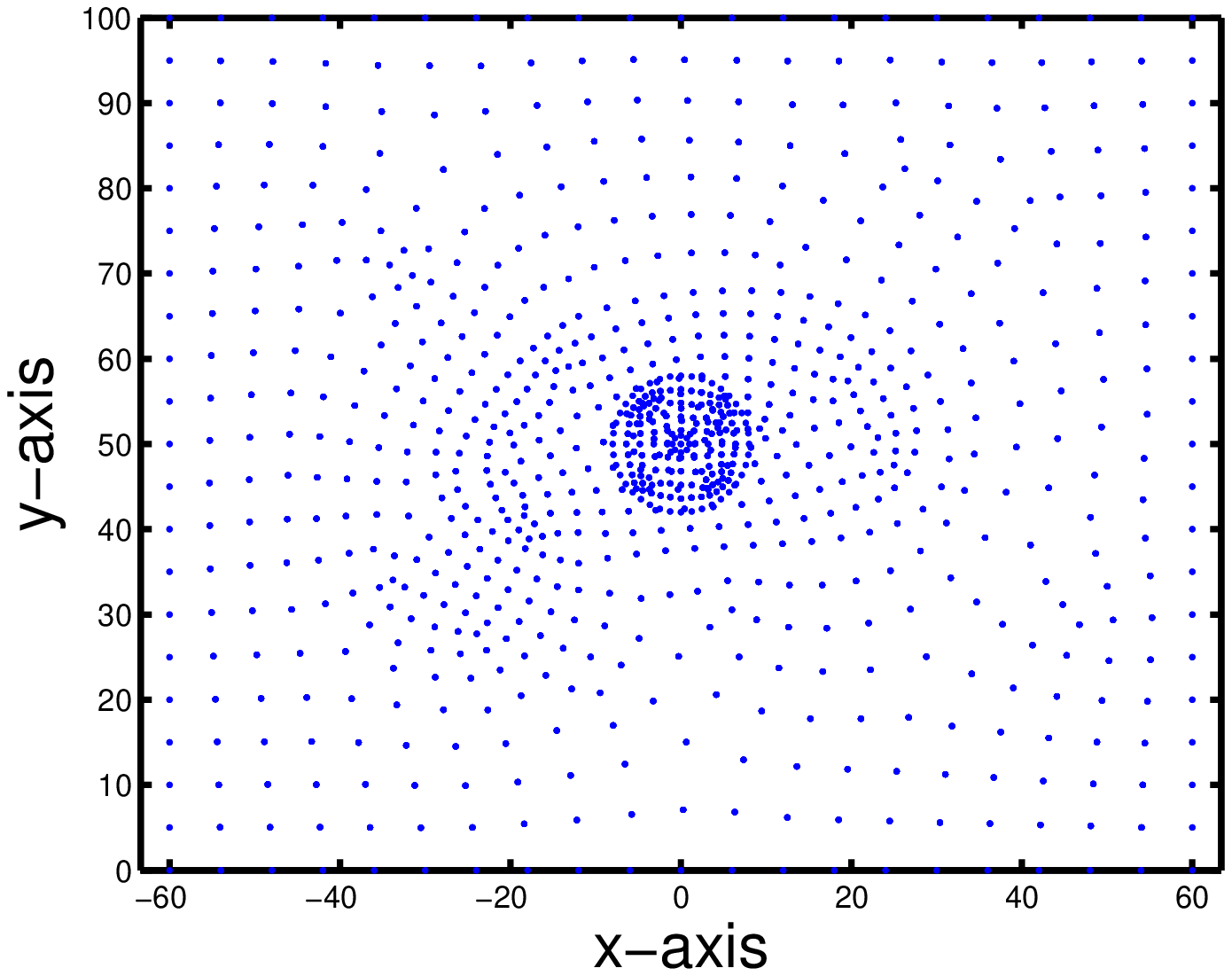}
\caption{~~(a)}
\end{subfigure}
\begin{subfigure}[b]{0.5\textwidth}
  \includegraphics[height=7cm,width=8.5cm]{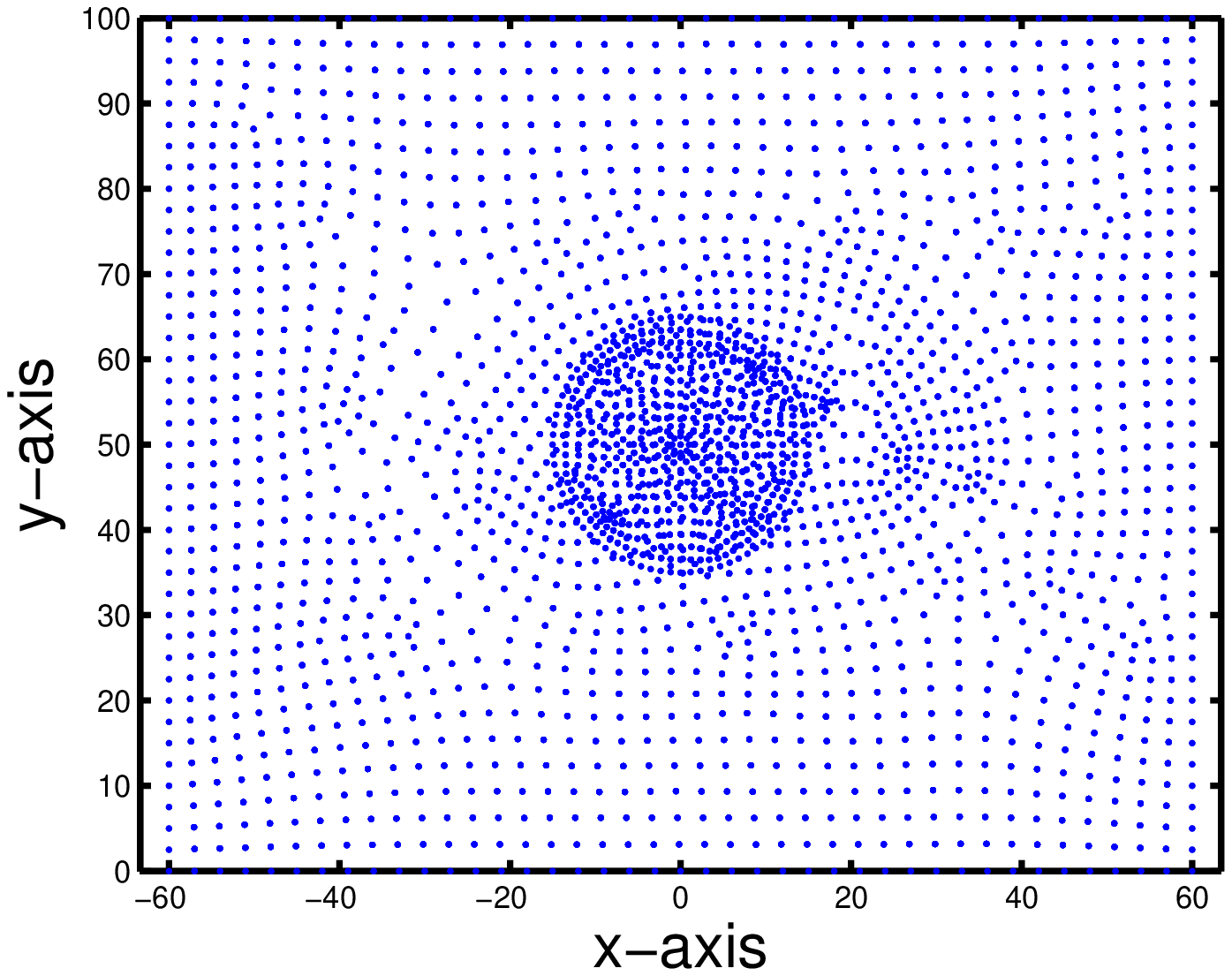}
\caption{~~(b)}
\end{subfigure}
\caption{Illustration of computational domain by nodal distribution (a) with coarse (754) nodes and (b) with finer (1968) nodes.}\label{fig2}
\end{figure}
 The unknown field variable $T(x, y)$ using MLS approximation is approximated by $T^{h}(x, y)$ over the $2D$ domain as details in \cite{L}.
\begin{align}
T(x,y)\approx T^{h}(x, y)=\sum^{m}_{j=1}{p_{j}(x,y)a_{j}(x,y)}\equiv \textbf{P}^{T}(x, y)\textbf{a}(x, y)\equiv \textbf{P}^{T}(\textbf{x})\textbf{a}(\textbf{x})\label{e15}
\end{align}
Where, $m$ is the number of terms in the basis, $a_{j}(x, y)$ is the non-constant coefficients, $p_{j}(x, y)$ is the monomial basis function, $\textbf{x}^{T}=[x ~y]$ and $\textbf{p}^{T}(\textbf{x})=[1 ~x ~y]$.\\
The unknown coefficients $a_{j}(\textbf{x})$ at any point $\textbf{x}$ are determined by minimizing the functional $J(\textbf{x})$ given by
\begin{align}
J(\textbf{x})=\sum^{n}_{I=1}{w(\textbf{x}-\textbf{x}_{I})}\left[\textbf{p}^{T}(\textbf{x})\textbf{a}(\textbf{x})-T_{I} \right]^{2}\label{e16}
\end{align}

Where $T_{I}$ is the nodal parameter at $\textbf{x}=\textbf{x}_{I}$ and these are not nodal values of $T^{h}(\textbf{x}=\textbf{x}_{I})$ being $T^{h}(\textbf{x})$ an approximate but not an interpolant; $w(\textbf{x}-\textbf{x}_{I})$ is a non-zero weight function over a small domain, called a support domain and n is the number of nodes in the domain of influence of $\textbf{x}$ for which $w(\textbf{x}-\textbf{x}_{I})\neq 0$.
The shape function $\Phi_{I}(\textbf{x})$ is defined as
\begin{align}
\Phi_{I}(\textbf{x})=\sum^{m}_{j=1}{p_{j}(\textbf{x})[A^{-1}(\textbf{x})\textbf{B}(\textbf{x})]_{jI}}=\textbf{p}^{T}\textbf{A}^{-1}\textbf{B}_{I} \label{e23}
\end{align}
Where $A$ and $B$ are defined as
\begin{align}
\textbf{A}(\textbf{x})&=\sum^{n}_{I=1}{w(\textbf{x}-\textbf{x}_{I})}\textbf{p}(\textbf{x}_{I})\textbf{p}^{T}(\textbf{x}_{I}) \label{e18}\\
\textbf{B}(\textbf{x})&=\left[{w(\textbf{x}-\textbf{x}_{1})}\textbf{p}(\textbf{x}_{1}),{w(\textbf{x}-\textbf{x}_{I})}\textbf{p}(\textbf{x}_{2}),...,{w(\textbf{x}-\textbf{x}_{n})}\textbf{p}(\textbf{x}_{n})\right]\label{e19}
\end{align}
\subsection{Description of Weight Function}
The choice of the weight function $w(\textbf{x}-\textbf{x}_{I})$ affects the resulting approximation $T^{h}(\textbf{x})$ in EFGM. In the EFGM, the smoothness as well as the continuity of the MLS approximations is governed by the smoothness and the continuity of the weight function $w(\textbf{x}-\textbf{x}_{I})$. Using cubic spline weight function in the current study, and normalized radius ($r$) is defined for:
\begin{align}
w(\textbf{x}-\textbf{x}_{I}) \equiv w(r)=
    \begin{cases}
            {\frac{2}{3}-4r^{2}+4r^{3}},                                           &          r\leq\frac{1}{2}\\
            {\frac{4}{3}-4r+4r^{2}-\frac{4}{3}r^{3}},                              &          \frac{1}{2}\leq r \leq {1} \label{e24}\\
            {0},                                                                   &          r>{1}
    \end{cases}
\end{align}

where $r=\frac{\|\textbf{x}-\textbf{x}_{I}\|}{d_{mI}}$, ${\|\textbf{x}-\textbf{x}_{I}\|}$ is the distance from an assessment point ($\textbf{x}$) to a node ($\textbf{x}_{I}$) and $d_{mI}$ is the size of the support domain at a particular node $I$. The size of the support domain is evaluated as $d_{mI}=d_{max}C_{I}$, where $d_{max}$ is the scaling parameter. The optimum range of the scaling parameter ($d_{max}$) is chosen in the interval $(1, 1.5)$. In present the problem, rectangular support domain is used because the rectangular shape one is more common for $2D$ space. The weight function at any point is obtained as
\par
$w(r)=w(r_{x})w(r_{y})$  where $r_{x}=\frac{|\textbf{x}-\textbf{x}_{I}|}{d_{mxI}}$,~$r_{y}=\frac{|\textbf{y}-\textbf{y}_{I}|}{d_{myI}}$, where $d_{mxI}=d_{max}c_{xI}$, $d_{myI}=d_{max}c_{yI}$,is the domain of influence of node $I$.
\subsection{Variational Formulation} The weighted integral form of the heat transfer equation ($\ref{e6}$) over the entire domain is obtained as
\begin{align}
\int \int_{\Omega} {w[(\rho c_{eff}) \frac{\partial T}{\partial t}-k(\frac{\partial^{2} T }{\partial x^{2}}+\frac{\partial^{2} T}{\partial y^{2}})-\omega_{b}(\rho_{b} c_{b}) (T_{b}-T)-Q_{m}+Q]}d\omega=0 \label{e25}
\end{align}
where, $w$ is arbitrary test function and may be viewed as the variation in $T$.
\subsection{Penalty Method for Imposition of Essential Boundary Conditions} Using the penalty method to impose boundary conditions the variational form ($\ref{e25}$) is written as
\begin{align}
\int \int_{\Omega} {w[(\rho c_{eff}) \frac{\partial T}{\partial t}-k(\frac{\partial^{2} T }{\partial x^{2}}+\frac{\partial^{2} T}{\partial y^{2}})-\omega_{b}(\rho_{b} c_{b}) (T_{b}-T)-Q_{m}+Q]}d\Omega+\int_{\Gamma}{\bar \alpha w (T-T_{\Gamma})d\Gamma}=0 \label{e26}
\end{align}
The weak form of equation ($\ref{e26}$) with boundary conditions is obtained as
\begin{align}
\int \int_{\Omega} {[w(\rho c_{eff}) \frac{\partial T}{\partial t}+k(\frac{\partial w }{\partial x}\frac{\partial T }{\partial x}+\frac{\partial w}{\partial y}\frac{\partial T}{\partial y})+\omega_{b}(\rho_{b} c_{b})w T]}d\Omega+\int_{\Gamma}{\bar \alpha w T d\Gamma} \nonumber\\
=(\omega_{b}(\rho_{b} c_{b}) T_{b}-Q_{m}+Q)\int \int_{\Omega}{w}d\Omega + \int_{\Gamma}{\bar \alpha w T_{\Gamma}d\Gamma}=0 \label{e26}
\end{align}
Where the test function($w$) is to be replaced by MLS shape function $\phi_{I}$ ($I=1,2,...N$), $N$ is the number of nodes in the whole domain. In this study, penalty parameter ($\bar \alpha$) is chosen as $10^{6}$. The solution of heat transfer equation ($\ref{e26}$), expression of stiffness matrix are obtained as:
\begin{align}
[M]\left\{\frac{\partial T}{\partial t}\right\}+[K]\left\{T\right\}=[F] \label{e27}
\end{align}
Where,
\begin{align}
M_{IJ}&=\rho c_{eff}\int \int_{\Omega}{\phi_{I}\phi_{J}d\Omega} \nonumber\\
K_{IJ}&=\int \int_{\Omega}\left[{k\left(\frac{\partial \phi_{I}}{\partial x}\frac{\partial \phi_{J}}{\partial x}+\frac{\partial \phi_{I}}{\partial y}\frac{\partial \phi_{J}}{\partial y}\right)+\omega_{b}(\rho_{b} c_{b})\phi_{I}\phi_{J}}\right]d\Omega+\bar\alpha\int_{\Gamma}{\phi_{I}\phi_{J}}d\Gamma \nonumber\\
F_{IJ}&=(\omega_{b}(\rho_{b} c_{b}) T_{b}-Q_{m}+Q)\int \int_{\Omega}{\phi_{I}}d\Omega + \int_{\Gamma}{\bar \alpha \phi_{I} T_{\Gamma}d\Gamma}\nonumber
\end{align}
 For time discretization, backward difference scheme is used. The equation (\ref{e27}) can be written as:
\begin{align}
[\tilde{K}+M][T]_{s+1}=[R_{s}]
\end{align}
where,
\begin{align}
R_{s}=[M+(1-\alpha){\Delta t}K]T_{s}+\Delta t F,~~~~ [\tilde{K}]=\alpha \Delta t K
\end{align}
\section{Results and Discussion}
\subsection{Validation and comparison}
The results obtained from EFGM are validated with $\alpha$-FEM results reported by Li et al. \cite{LLTH}.  In this study, we have considered a circular probe of radius $1mm$ for cooling purpose. The temperature variable kept at the target site of tumor center reported by Li et al. \cite{LLTH} by means of $\alpha$-FEM techniques are evaluated against present EFGM results as shown in fig.[3] following $600$ sec of freezing and the results obtained are observed to lie well within the acceptable range.

\begin{figure}[!b]
  \begin{minipage}[!b]{0.5\linewidth}
  \centering
  \includegraphics[height=6.5cm,width=8cm]{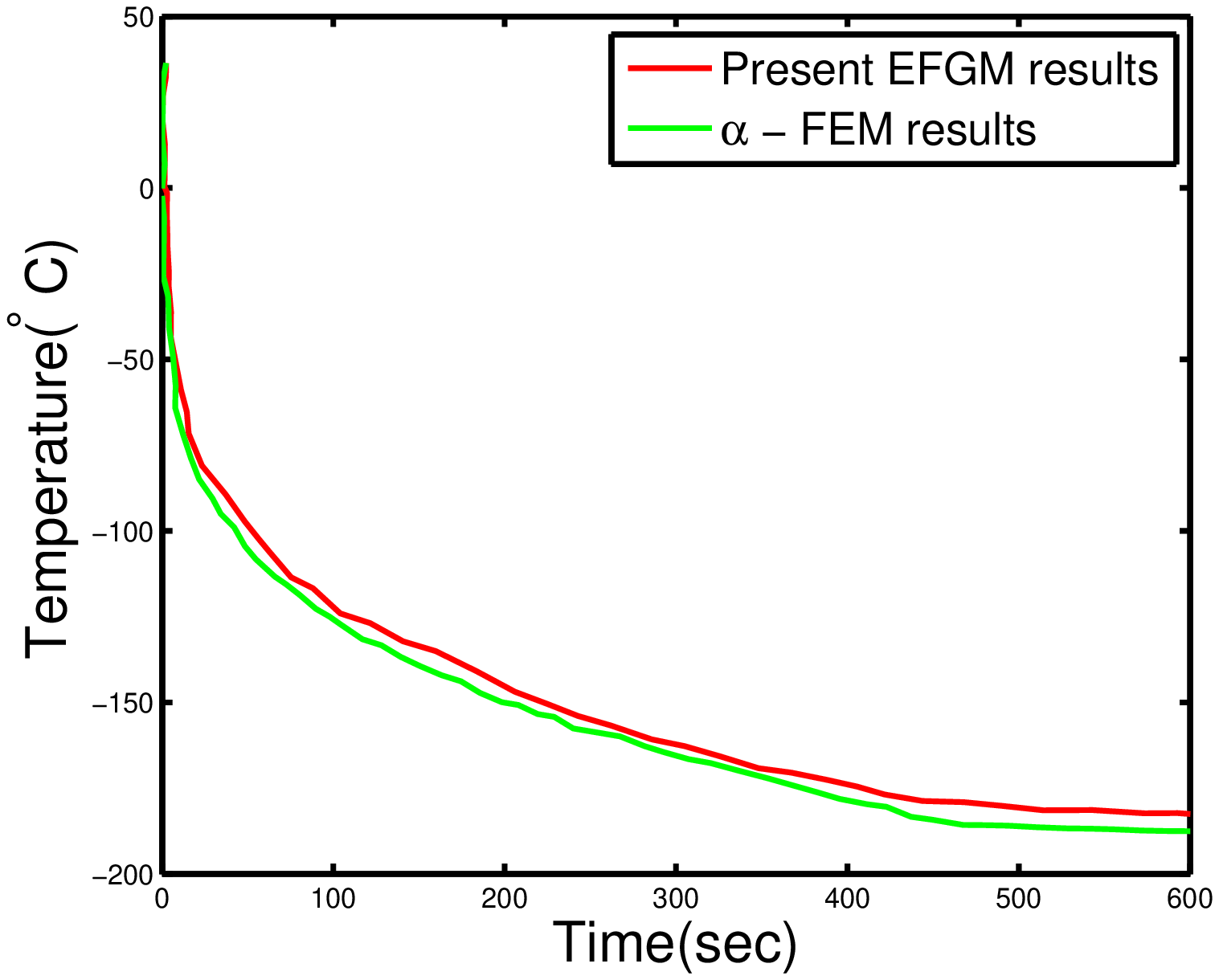}
  \vspace{-0.1in}
  \caption{Comparison of temperature distribution at the tumor center with results provided by Li et al. \cite{LLTH}.}
  \label{fig3}
  \end{minipage}
  \hspace{0.9cm}
  \begin{minipage}[!b]{0.5\linewidth}
  \includegraphics[height=6.5cm,width=8cm]{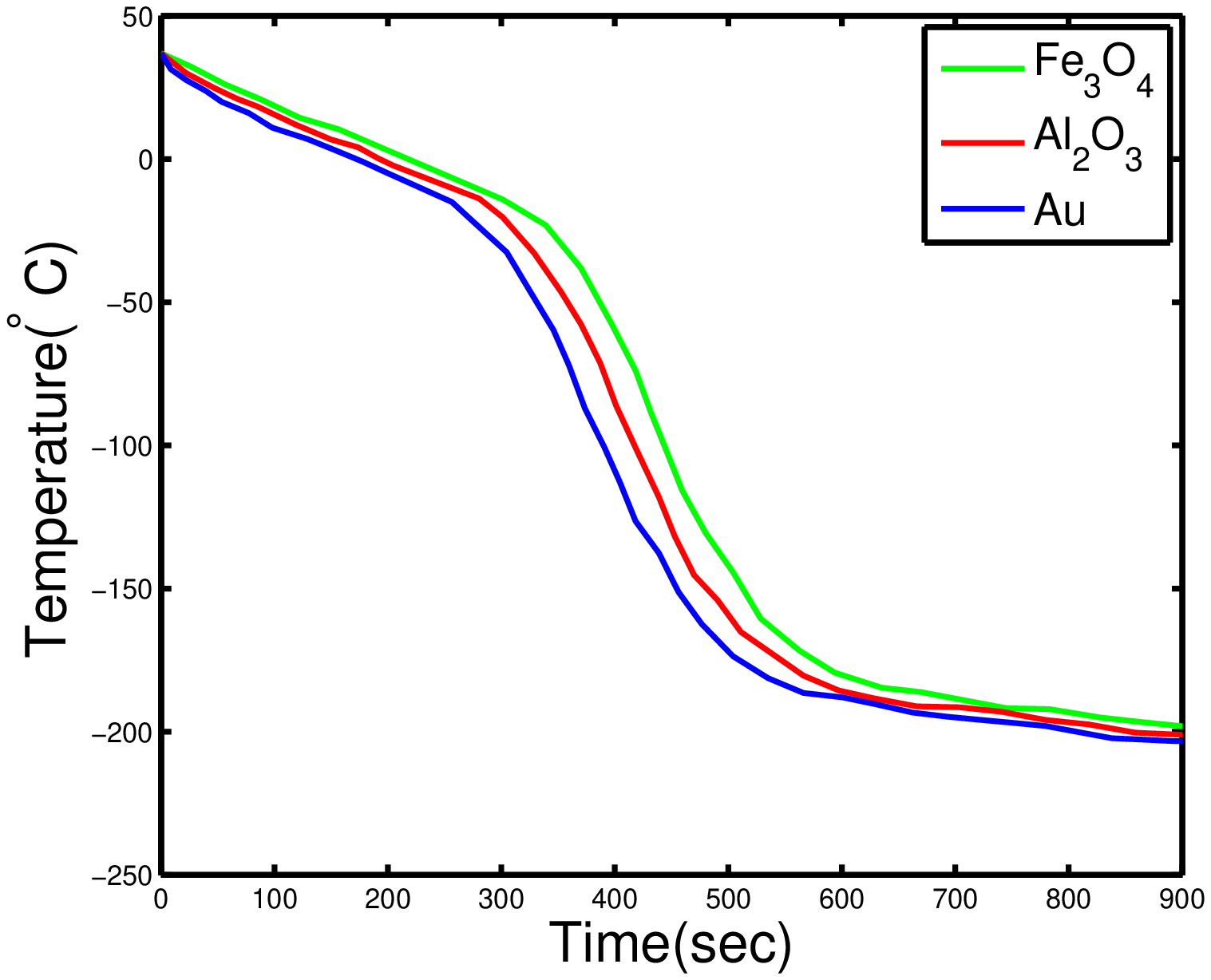}
  \caption{Temperature distribution with time at the tumor center for different nanoparticles with various parameters $\phi=0.20$, $d_{p}=10 nm$ and $\delta=20$.}
  \label{fig4}
  \end{minipage}
\end{figure}

\subsection{Discussion}
The temperature field variable determined by using EFGM is shown in fig. 3 to 8. Table 1 contains the physical and thermal property of loaded nanoparticles and biological tissue. The single probe is used for freezing process. The probe functions as a heat sink and maintains freezing at a constant rate of $7.2\times10^{-7} W/m^{3} s$ at the targeted tumor (irregular) tissue.\\
In all the cases, the default value of the parameters $d_{p}$ (diameter of the nanoparticle), $\phi$ (concentration of nanoparticles) and $\delta$ (ratio of base fluid conductivity to nanolayer) were chosen as $10nm$, $20\%$ and $20$ respectively. For numerical simulation different number of nodes is distributed in the given computational domain according to the probe or our requirement. Initially, we have considered 754 numbers of nodes, when size of ice balls were enhanced with time. Four point Gauss quadrature is used for numerical integration.

\vspace{2cm}

\textbf{Table 1.}\\
Physical and thermal properties of biological tissue and different nanoparticles \cite{YL,LLTH}.
\begin{center}
\begin{tabular}{l|l|l|l}
 \hline
   Items                                       & Symbols    & Units                     & Values               \\
                                               &            &                           &                      \\
\hline
 Thermal conductivity of the frozen tissue     & $k_f$      & $W/m$ $ ^{0}C$            & $2.0$               \\
 Thermal conductivity of the unfrozen tissue   & $k_u$      & $W/m $ $^{0}C$            & $0.5$                \\
 Specific heat capacity of the frozen tissue   & $c_f$      & $J/m^{3} $ $^{0}C$        & $1.8\times10^{3}$    \\
 Specific heat capacity of the unfrozen tissue & $c_u$      & $J/m^{3}$ $ ^{0}C$        & $3.6\times10^{3}$    \\
 Latent heat                                   & $\bar h$   & $J/Kg$                    & $4.2\times10^{6}$    \\
 Temperature of lower phase change             & $T_{s}$    & $^{0}C$                   & $-8$                 \\
 Temperature of upper phase change             & $T_{l}$    & $^{0}C$                   & $-1$                 \\
 Metabolic rate of the liver                   & $Q_{m}$    & $W/m^{3}$                 & $4200$               \\
 Blood perfusion rate                          & $w_{b}$    & $ml/s/ml$                 & $0.0005$             \\
 Density of the frozen tissue                  & $\rho_f$   & $Kg/m^{3}$                & $1000$                \\
 Density of the unfrozen tissue                & $\rho_u$   & $Kg/m^{3}$                & $1000$                \\
 Density of blood                              & $\rho_b$   & $Kg/m^{3}$                & $1000$                \\
 Thermal conductivity of $Fe_3O_4$             & $k_p$      & $W/m$ $^{0}C$             & $7.1$                 \\
 Specific heat capacity of $Fe_3O_4$           & $c_p$      & $J/m^{3}$ $ ^{0}C$        & $3.2\times10^{3}$      \\
 Density of $Fe_3O_4$                          & $\rho_p$   & $Kg/m^{3}$                & $4800$                 \\
 Thermal conductivity of $Al_2O_3$             & $k_p$      & $W/m$ $^{0}C$             & $39.7$                 \\
 Specific heat capacity of $Al_2O_3$           & $c_p$      & $J/m^{3}$ $^{0}C$         & $2.82\times10^{3}$     \\
 Density of $Al_2O_3$                          & $\rho_p$   & $Kg/m^{3}$                & $3970$                 \\
 Thermal conductivity of $Au$                  & $k_p$      & $W/m$ $^{0}C$             & $297.7$                \\
 Specific heat capacity of $Au$                & $c_p$      & $J/m^{3}$ $ ^{0}C$        & $2.21\times10^{3}$     \\
 Density of $Au$                               & $\rho_p$   & $Kg/m^{3}$                & $19320$                \\
 \hline
\end{tabular}
\end{center}

\textbf{Table 2.}\\
 Temperature calculated at the tumor centre and different point at the tumor tissue, after 15 min of freezing for different nanoparticles ($d_p=10nm,~~\phi=0.20,~~\delta=20$).
\begin{center}
.\small
\begin{tabular}{l|l|l|l}
 \hline
   Nano                 & Value of temperature &Minimum value of Temperature   &Minimum value of Temperature      \\
  particle              & at a point(0,50)     &at a point (0,64), 14 mm apart &at tumor boundary,at point (0,68) \\
                        & (i.e.,tumor center)  &from the tumor centre          & 18 mm apart from the tumor centre \\
\hline
  $Fe_{3}O_{4}$         & -198                 &-41.3313                       &-15.2222                          \\
  $Al_{2}O_{3}$         & -202                 &-42.6667                       &-16.1111                           \\
  $Au$                  & -205                 &-43.6667                       &-16.7778                           \\
 \hline
\end{tabular}
\end{center}
\begin{figure}[!b]
\begin{subfigure}[b]{0.5\textwidth}
  \includegraphics[height=5cm,width=7cm]{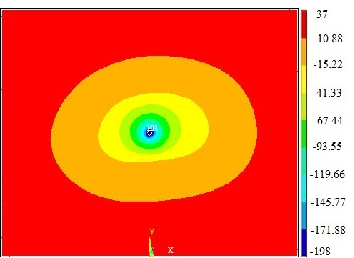}
\caption{(a)}
\end{subfigure}
\hspace{3.0mm}
\begin{subfigure}[b]{0.5\textwidth}
  \includegraphics[height=5cm,width=7cm]{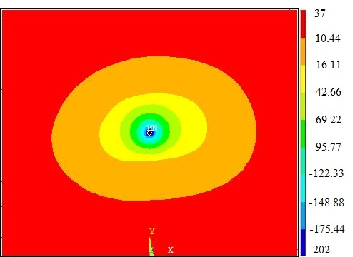}
\caption{(b)}
\end{subfigure}
\vspace{2.0mm}
\begin{subfigure}[b]{0.5\textwidth}
  \includegraphics[height=5cm,width=7cm]{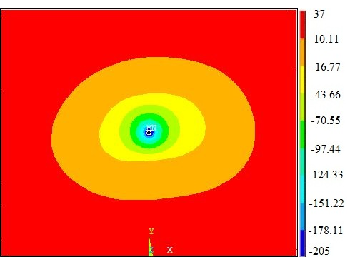}
\caption{(c)}
\end{subfigure}
\hspace{3.0mm}
\begin{subfigure}[b]{0.5\textwidth}
  \includegraphics[height=5cm,width=7cm]{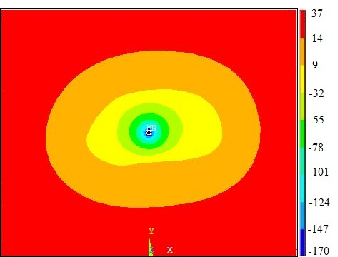}
\caption{(d)}
\end{subfigure}
\caption{Effect of different nanoparticles on cooling rate after $15$ min of freezing with (a) $Fe_3O_4$ (magnetite) nanoparticle (b) $Al_2O_4$ (alumina) nanoparticle (c) $Au$ (gold) nanoparticle, and (d) without nanoparticle.}\label{fig5}
\end{figure}

Temperature distribution at the tumor centre is shown in fig. 4 with different nanoparticles gold ($Au$), alumina ($Al_{2}O_{3}$) and magnetite ($Fe_{3}O_{4}$) as determined during $15$ min of freezing process. It is observed that various types of nanoparticles with different concentrations have variable effect on the cell due varied cooling rate. Fraction of larger volume of nanoparticles encompassing high thermal conductivity may possibly enhance the cooling rate at the target cell. Table 1 also contains heat capacity as well as thermal conductivity properties of different nanoparticles.\\
The values of temperature determined for two different points at the tumor tissue having distances $14.0$ mm (i.e. at point $(0, 64))$ and $18.0$ mm from the centre of tumor, after $15$ min of freezing with various nanoparticles and temperature responses are shown in table 2. It is quite clear that the temperature profiles are almost same at the tumor center. When using the nanoparticle, temperature profile decreases away from the tumor centre due to increased cooling efficiency.\\
Effect of different nanoparticles such as magnetite ($Fe_3O_4$), alumina ($Al_2O_3$) and gold ($Au$) with respect to variable  temperature profile at the targeted tumor tissues after $15$ min of freezing are shown in figures 5(a)-5(c). It is quite evident that the freezing system is changed in the presence of nanoparticles. In figure 5(a), for magnetic ($Fe_3O_4$) nanoparticles, the lowest obtained value of temperature was observed at $14 mm$ away from the tumor centre i.e. at point $(0,64)$) is $-41.33$ $^{0} C$. Similarly, for alumina nanoparticles ($Al_2O_3$), the value of minimum temperature is observed to be $-42.66$ $^{0}C$, shown in fig. 5(b). As in figure 5(c), for gold ($Au$) nanoparticles, this value is obtained $-43.66$ $^{0} C$. It appears as metal nanoparticles have high thermal conductivity, which increases heat conduction when they are injected into the tissue. Gold nanoparticles compared to others have the highest thermal conductivity. So under same freezing conditions, gold nanoparticles were found to manifest minimum value of temperature at the selected portion. Furthermore, shape, size and growth of ice-ball, are significantly affected by nanoparticles. In figures 5(a), 5(b) and 5(c), the temperature distributions are observed between $-41 ^{0}C$ to $-44 ^{0}C$ for time $15$ min with nanoparticles whereas, in the figure 5(d) the temperature distribution is found $-32 ^{0}C$ during 15 min without nanoparticle. Hence, nanoparticles are found to give a better freezing response in compare to normal fluids.
\begin{figure}[!b]
\begin{subfigure}[b]{0.5\textwidth}
  \includegraphics[height=6.5cm,width=8.5cm]{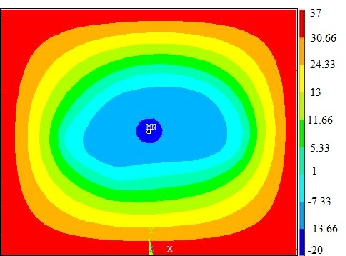}
\caption{~~(a)}
\end{subfigure}
\hspace{4.0mm}
\begin{subfigure}[b]{0.5\textwidth}
  \includegraphics[height=6.5cm,width=8.5cm]{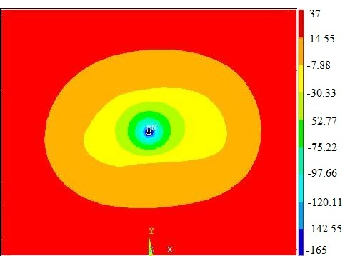}
\caption{~~(b)}
\end{subfigure}
\caption{Temperature distribution within the part of the human body with $15\%$ loaded of nanoparticles as $Al_{2}O_{3}$ (a) during 5 min and (b) during 10 min}
\end{figure}
The temperature contours of the freezing process are demonstrated in figures 6(a) and 6(b) after $5min$ and $10min$ respectively. The concentration of nanoparticles is considered $15 \%$ in which alumina nanoparticles are loaded in tumor tissue. It is found that the initial temperature ($37 ^{0}C$) decreases with increasing time. At $t=5 min$, the temperature at the tumor centre is obtained $-20 ^{0} C$. Here initial phase change occurs from liquid to solid regime. After freezing of $10 min$, the temperature at the tumor centre is observed $-165 ^{0}C$ while temperature of the periphery of $8 mm$ circular region around the tumor centre is $-52.66 ^{0}C$, which lies within $-40 ^{0}C$ to $-60 ^{0}C$ (i.e critical temperature) for destroying the tumor tissue. It is found that temperature is decreasing in the domain of tumor tissue and greater than $15 \%$ area successfully covered during $10 min$, without loss of tumor tissue.

\begin{figure*}[h!]
  \begin{minipage}[!b]{0.5\linewidth}
  \centering
  \includegraphics[height=7cm,width=8.5cm]{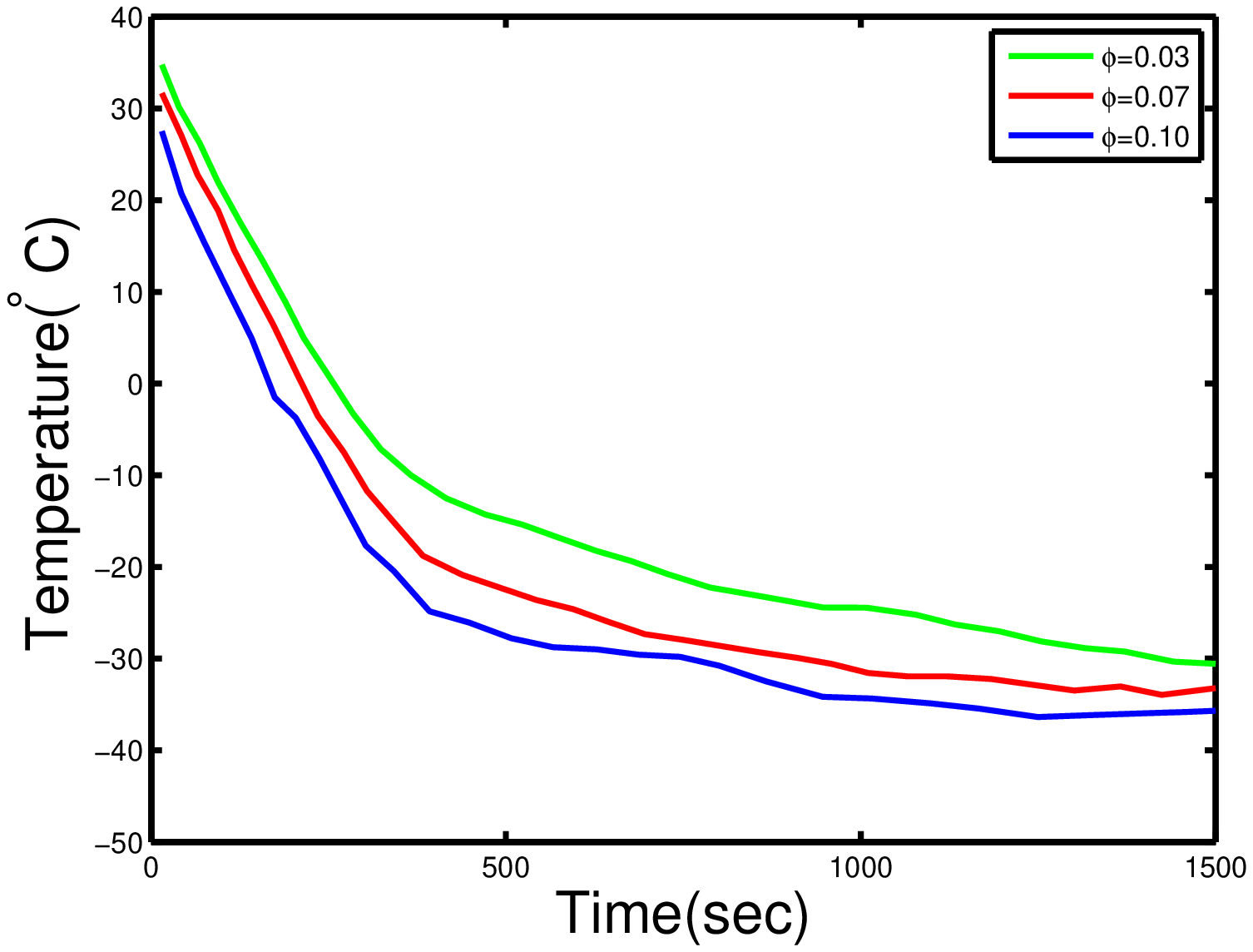}
  \vspace{-0.3in}
  \caption{Temperature distribution for time $25$ min at distance $18$ mm from the tumor centre for different concentration of alumina nanoparticles ($Al_2O_3$) with $d_p=10 nm$ and $\delta=20$.}
  \label{fig3}
  \end{minipage}
  \hspace{0.9cm}
  \begin{minipage}[!b]{0.5\linewidth}
  \includegraphics[height=7cm,width=8.5cm]{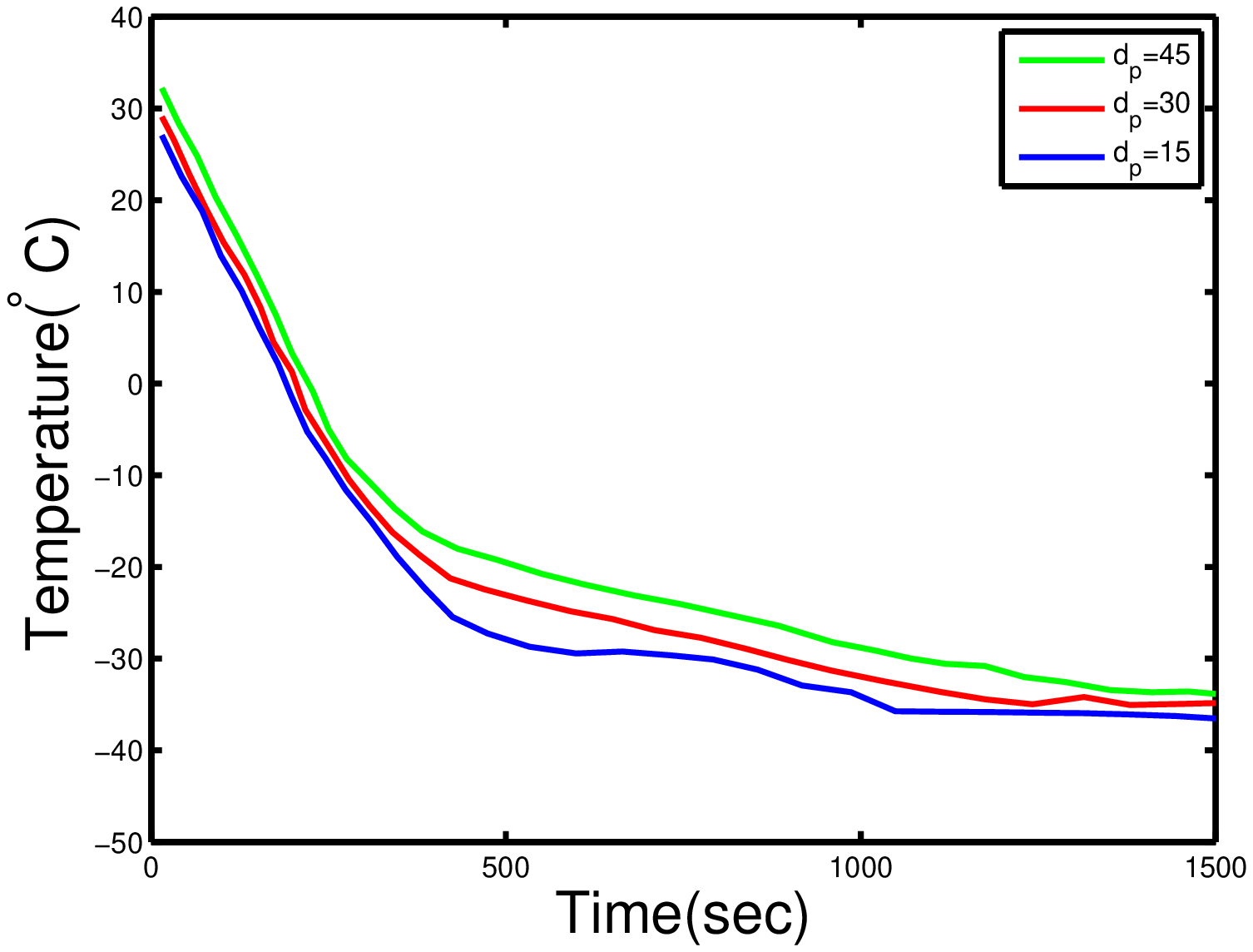}
 \vspace{-0.3in}
  \caption{Temperature distribution for time $25$ min at distance $18$ mm from the tumor centre for different diameter of alumina nanoparticle ($Al_2O_3$) with $\phi=0.20$ and $\delta=20$.}
  \label{fig4}
  \end{minipage}
\end{figure*}

\begin{figure}[h!]
\vspace{-0.10in}
\centerline{\includegraphics[height=7.0cm,width=8.0cm]{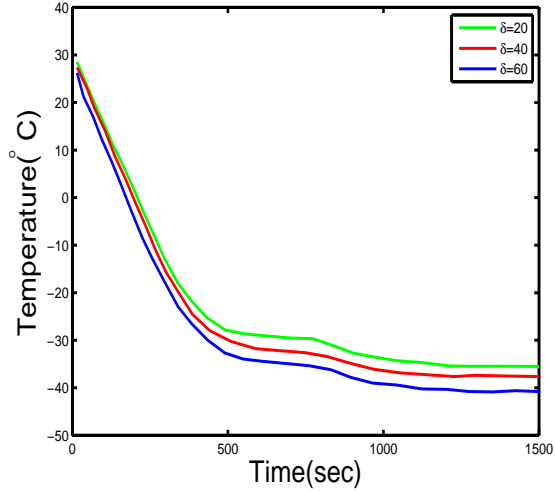}}
\vspace{-0.12in}
\caption{Temperature distribution for time $25$ min at distance $18$ mm from the tumor centre for different nanolayer to base fluid conductivity ratio with $d_{p}=10 nm$ and $\phi=0.20$.} \label{fig9}
\end{figure}
\par
The effect of nanoparticle concentration on the temperature profile is shown in Figure 7. We can clearly observe that large nanoparticle concentration and high thermal conductivity of can certainly enhance the cooling rate at the target cell. Temperature distributions by $18$ mm from the tumor centre are found with various concentrations of alumina ($Al_2O_3$) nanoparticles after $25$ min of freezing. It is also found that temperature profile decreases (i.e freezing rate increases) with increasing volume fraction of nanoparticles. Also we have found that greater than $45 \%$ region around the tumor centre is successfully covered by a single probe during $25$ min of cooling with the value of concentration ($\phi$) from $3 \%$ to $10 \%$.
\par
Variations in temperature distribution for different sizes of nanoparticles at a distance of $18$ mm from the tumor centre are shown in figure 8. For alumina ($Al_{2}O_{3}$), different size nanoparticles having diameter ranging from $1 \leq d_{p}\leq 45$ are taken. Cooling rate was enhanced by using nanoparticles of larger diameter as reported earlier. In similar way, the temperature profiles can be enhanced by enhancing increasing the diameter as capacity of thermal conductivity from nanoparticles to the base fluid is reduced. The capacity of bigger nanoparticles is lower than that of smaller particles.
\par
The effect of nanolayer to base fluid conductivity ratio ($\delta$) on temperature profiles (i.e. freezing rate) for alumina ($Al_2O_3$) nanoparticles are given in figure 9. The cylindrical nanoparticle has an interfacial layer. Orderness $\delta$($>1$) of the fluid particles at the interface depends on the, nature and surface chemistry of nanofluid. For alumina nanoparticle, the value of $(\delta)$ nanolayer to base fluid conductivity ratio lies from $1$ to $65.2529$ $(1< \delta < \frac{kp}{kf})$. Temperature is observed to be reduced with increment in nanolayer to base fluid conductivity ratio $(\delta)$ due to enhanced thermal conductivity. Finally, we have observed that for the value of d from $20$ to $60$, the detected region around the tumor centre was more than $45 \%$, during $25$ min of freezing rate with minimum loss of desired tissue.
\section{Conclusions}
Following important conclusions are observed.
\begin{itemize}
\item Maximum cooling efficiency of the tumor tissue is obtained in presence of nanoparticles of alumina ($Al_2O_3$), Iron ($Fe_3O_{4}$) and gold ($Au$). We have also observed that the gold nanoparticles have the maximum thermal conductivity.
\item Besides enhancing the cooling rate, the shape, size and growth of ice-crystals are controlled during cooling in the presence of nanoparticles.
\item It has been observed that greater than ($15 \%$) area (i.e. malignant tissue region) covered by single probe without loss of desired tissue during $10$ min, but after $25$ min, we have obtained greater than ($45 \%$) region (i.e. malignant tissue region) covered by the single probe but a minimum loss of desired tissue.
\item Murshed model with cylindrical nanoparticles is sufficiently good for freezing purpose as compared to other model (i.e. leong model with spherical nanoparticle).
\item Freezing efficiency, necessary to destroy the malignant tumor, increases with increasing concentration of nanoparticle and nanolayer to base fluid conductivity ratio and decreases with increasing the nanoparticle diameter.
\end{itemize}
\textbf{Acknowledgments}\\
Authors gratefully acknowledge for the financial support from SERB, New Delhi sponsored through a project no. SR/S4/MS:713/10.

\end{document}